\documentclass[twocolumn, tighten]{aastex63}
\usepackage{natbib}

\newcommand{\bo}{JVAS~B0218+357}

\received{2020 July 10}
\accepted{2020 July 28}
\shorttitle{Millimeter-VLBI observations of B0218+357}
\shortauthors{Hada et al.}
\begin{document}

\title{Millimeter-VLBI Detection and Imaging of the Gravitationally Lensed $\gamma$-Ray Blazar JVAS B0218+357}

\correspondingauthor{Kazuhiro Hada}
\email{kazuhiro.hada@nao.ac.jp}

\author{Kazuhiro Hada}
\affil{Mizusawa VLBI Observatory, National Astronomical Observatory of Japan, 2-12 Hoshigaoka, Mizusawa, Oshu, Iwate 023-0861, Japan}
\affil{Department of Astronomical Science, The Graduate University for
Advanced Studies (SOKENDAI), 2-21-1 Osawa, Mitaka, Tokyo 181-8588, Japan}

\author{Kotaro Niinuma}
\affil{Graduate School of Sciences and Technology for Innovation, Yamaguchi University, Yoshida 1677-1,
Yamaguchi, Yamaguchi 753-8512, Japan}

\author{Julian Sitarek}
\affil{University of \L\'{o}d\'{z}, 90236 \L\'{o}d\'{z}, Poland}

\author{Cristiana Spingola}
\affil{INAF$-$Istituto di Radioastronomia, via Gobetti 101, I-40129 Bologna, Italy}
\affil{Dipartimento di Fisica e Astronomia, Universit\`a degli Studi di Bologna, via Gobetti 93/2, I-40129 Bologna, Italy}

\author{Ayumi Hirano}
\affil{Graduate School of Sciences and Technology for Innovation, Yamaguchi University, Yoshida 1677-1,
Yamaguchi, Yamaguchi 753-8512, Japan}

\begin{abstract}
We observed the gravitationally lensed blazar \bo{} with the KVN and VERA Array (KaVA) at 22, 43, and 86\,GHz. The source has recently been identified as an active $\gamma$-ray source up to GeV/TeV energy bands, rendering a unique target for studying relativistic jets through gravitational lensing. Here we report the first robust VLBI detection and imaging of the lensed images up to 86\,GHz. The detected mas-scale/parsec-scale morphology of the individual lensed images (A and B) is consistent with that previously seen at 22 and 15\,GHz, showing the core--jet morphology with the jet direction being the same as at the low frequencies. The radio spectral energy distributions of the lensed images become steeper at higher frequencies, indicating that the innermost jet regions become optically thin to synchrotron emission. Our findings confirm that the absorption effects due to the intervening lensing galaxy become negligible at millimeter wavelengths. These results indicate that high-frequency VLBI observations are a powerful tool to better recover the intrinsic properties of lensed active galactic nucleus jets, which therefore allow us to study the interplay between the low- and high-energy emission.

\end{abstract}




\section{Introduction} \label{sec:intro}

The gravitational lensing (GL) effect can
occur when a foreground massive object, such as a galaxy, lies close to the line of sight to a background source~\citep[e.g.,][]{treu2010, congdon2018}. As a consequence of GL the background source may be distorted and magnified into multiple lensed images allowing us to reveal objects that would be otherwise impossible to detect with current facilities~\citep{hartley2019}. Observations of GL systems can be used for a wide variety of astrophysical applications, from inferring the properties of the deflectors~\citep[e.g.,][]{gilman2020} to the detailed study of the most distant galaxies~\citep[e.g.,][]{vanzella2020}. If the background source is radio-loud it is also possible to overcome obscuration caused by dust and microlensing effects due to stars in the lensing galaxy (e.g., \citealt{koopmans2003}, but see also \citealt{koopmans2000}).
Moreover, radio-loud, strongly lensed sources can be observed with interferometric arrays at angular resolution from milliarcsecond (mas) at centimeter wavelengths to sub-mas scales when observing at millimeter (mm) wavelengths. Interferometers are capable of monitoring observations that can be used to measure cosmological parameters  \citep[e.g.,][]{refsdal1964} and proper motions at high redshifts (e.g., \citealp{spingola2019}). High-resolution radio observations can also provide precise measurements of the lensed images positions and flux densities, which are essential to constrain lens mass models.

JVAS\footnote{Jodrell Bank VLA Astrometric Survey.}~B0218+357 is a radio-loud gravitationally lensed active galactic nucleus (AGN) originally identified by \citet{patnaik1993}. The background source is located at a redshift of $z=0.944$~\citep{cohen2003} and lensed by a foreground spiral galaxy B0218+357G at $z=0.685$~\citep{browne1993}. The lensing effect splits the AGN into two lensed images, A and B, separated by $\sim$335\,mas, but also distorts the background source in a bright and complete Einstein ring at radio~\citep{odea1992, patnaik1993, biggs2003, wucknitz2004}. Radio monitoring observations of the lensed images A and B inferred a time delay of 10--12 days, where the lensed image A is confirmed as the leading component~\citep{corbett1996, biggs1999, cohen2000, biggs2018}. 

In recent years \bo{} has attracted a great deal of interest from the community of high-energy astrophysics. In 2012 August the Large-Area-Telescope (LAT) detector on board the \textit{Fermi} $\gamma$-ray satellite detected bright $\gamma$-ray flares from this source~\citep{cheung2014}. Although the instrument does not have enough angular resolution to resolve A and B, an autocorrelation analysis of the $\gamma$-ray light curves derived a time delay of $11.46\pm0.16$ days, which is consistent with a recently updated time delay of $11.3\pm0.2$ days in the radio band~\citep{biggs2018}. \textit{Fermi}-LAT detected another $\gamma$-ray flare from the source in 2014 July~\citep{buson2015}. For this event, the source was further detected in very-high-energy (VHE; $\gtrsim$100\,GeV) $\gamma$-rays with the MAGIC~\citep{ahnen2016}, which set a new record of the most distant VHE $\gamma$-ray-detected source in the universe. Since GL may offer a powerful tool to probe the site of $\gamma$-ray emission in a relativistic jet~\citep{neronov2015, barnacka2016, sitarek2016, vovk2016}, observations of \bo{} together with high-resolution instruments at other wavelengths are of great importance.  

At radio, the parsec(pc)-scale structures of the lensed system were resolved by very long baseline interferometry (VLBI) observations. Both A and B showed core-jet morphology, typical of powerful blazars~\citep{biggs2003, mittal2006, spingola2016}. The observed flux magnification ratio A/B was typically 3--4 between 8 and 22\,GHz. However, \citet{mittal2006} found the frequency-dependent nature of the lensed (A and B) images, their radio spectra and flux ratios, which may be most likely explained by absorption effects in the foreground lens galaxy~\citep{mittal2007}. In fact, the A/B flux ratio is inverted in optical bands due to the strong extinction toward A~\citep{falco1999}.

Nevertheless, the source has hardly been investigated at mm wavelengths such as 43\,GHz (7\,mm) and 86\,GHz (3.5\,mm). While there was an early VLBI experiment at 43\,GHz~\citep{porcas1996} and a snapshot flux measurement at 86\,GHz~\citep{lee2008}, these reports were inadequate to quantify the properties of the source at these frequencies. Probing this source at high frequencies may provide insight into the nature of the innermost jet regions of a lensed system and help understand the origin to the high-energy activities. Here we report the first results of high-frequency VLBI observations of \bo{} obtained with KaVA~\citep{niinuma2014}, a joint array of the Korean VLBI Network \citep[KVN;][]{lee2014} and the VLBI Exploration of Radio Astrometry \citep[VERA;][]{kobayashi2003}. These observations were conducted as part of large multiwavelength (MWL) campaigns led by the MAGIC Collaboration, and a detailed study based on the broadband MWL data will be presented in a separate publication (MAGIC Collaboration et al. in prep.). In the next section we describe our radio observations and data reduction. In Sections 3 and 4, our results are presented and discussed. In the final section we summarize the paper. Throughout this paper, we assume $H_0=67.8\; \mathrm{km\,s^{-1}~Mpc^{-1}}$, $\Omega_{\rm M}=0.31$, and $\Omega_{\Lambda}=0.69$ \citep{planck2018}.

\section{Observations and Data Reduction} \label{sec:obs}

\subsection{KaVA 22 and 43\,GHz}
Between 2017 and 2018 January, we observed \bo{} with KaVA for 9 sessions as part of coordinated MWL campaigns on this source. Each session lasted 2 consecutive nights, where we had a 5--8-hour track at 22\,GHz for the first day and another similar track at 43\,GHz for the following day. By default 7 stations (3 from KVN and 4 from VERA) joined each session, but occasionally VERA-Mizusawa or VERA-Ishigaki was missing due to local issues (see Table~\ref{tab:observations}). All the data were recorded at 1\,Gbps (a total bandwidth of 256\,MHz with eight 32-MHz subbands). Left-hand circular polarization was received and the data were correlated by the Daejeon hardware correlator~\citep{lee2015}. The initial data calibration (amplitude, phase, bandpass) was performed using the National Radio Astronomy Observatory (NRAO) Astronomical Image Processing System \citep[AIPS;][]{greisen2003} based on the standard KaVA/VLBI data reduction procedures~\citep{niinuma2014, hada2017, park2019}. The data were finally averaged in frequency, but individual subbands were kept separate to avoid bandwidth smearing. Similarly, the data were time-averaged to only 16 seconds to avoid time smearing. Following the initial calibration, imaging was performed in the Difmap software~\citep{shepherd1994} using the standard CLEAN and self-calibration procedures. The visibility amplitude of KaVA data was self-calibrated down to a solution interval of 10--120 minutes.

In this paper, we present the KaVA results for 4 out of the 9 sessions where we had simultaneous KVN 86\,GHz data (see below). Our regular monitoring of \bo{} with KaVA was continued until early 2019 and detailed analyses of the structural evolution and time-domain properties using the whole KaVA datasets will be presented in future papers.

\begin{table}[ttt]
 \begin{minipage}[t]{1.0\columnwidth}
  \centering 
  \caption{KaVA and KVN observations of \bo{}\label{tab:observations}}
    \footnotesize
    \begin{tabular*}{1.0\columnwidth}{@{\extracolsep{\fill}}lccccl}
    \hline
    \hline 
     Date & Band & Array & Beam size & $S_{\rm tot}$ & $I_{\rm rms}$ \\
          & (GHz)     &          & (mas$\times$mas, deg) & (mJy) & $\left(\frac{\rm mJy}{\rm bm}\right)$ \\
%          & (a)       & (b)      & (c)                     & (d)    \\
    \hline
     Jan/14/17 & 22 & KaVA$^{*}$ & 1.63$\times$1.27, $49$ & 408 & 0.25 \\
     Jan/15/17 & 43 & KaVA$^{*}$ & 1.06$\times$0.79, $63$ & 292 & 0.31 \\
     Jan/15/17 & 86 & KVN & 1.35$\times$0.76, $89$ & 172 & 1.05 \\
    \hline
     Oct/16/17 & 22 & KaVA & 1.17$\times$1.11, $6$ & 401 & 0.47 \\
     Oct/17/17 & 43 & KaVA$^{**}$ & 0.58$\times$0.55, $18$ & 299 &  0.51  \\
     Oct/17/17 & 86 & KVN  & 1.42$\times$0.76, $-72$ & 103 & 1.06 \\
    \hline
     Nov/11/17 & 22 & KaVA & 1.52$\times$1.19, $14$ & 416 & 0.38 \\
     Nov/12/17 & 43 & KaVA$^{**}$ & 0.88$\times$0.58, $24$ & 238 &  0.31 \\
     Nov/12/17 & 86 & KVN & 1.36$\times$0.75, $-79$ & 84 & 0.72 \\
    \hline
     Jan/4/18 & 22 & KaVA & 1.31$\times$1.11, $4$ & 401 & 0.33 \\
     Jan/5/18 & 43 & KaVA & 0.69$\times$0.55, $-32$ & 298 & 0.59 \\
     Jan/5/18 & 86 & KVN & 1.35$\times$0.75, $-83$ & 102 & 0.67 \\
    \hline
    \end{tabular*}
  \end{minipage}
  \tablecomments{*: KaVA without VERA-Mizusawa. **: KaVA without VERA-Ishigaki. $S_{\rm tot}$: VLBI total (A+B) flux density. $I_{\rm rms}$: off-source 1$\sigma$ image rms noise level.} 
\end{table}

\begin{figure}[ttt]
 \centering
 \includegraphics[angle=0,width=0.98\columnwidth]{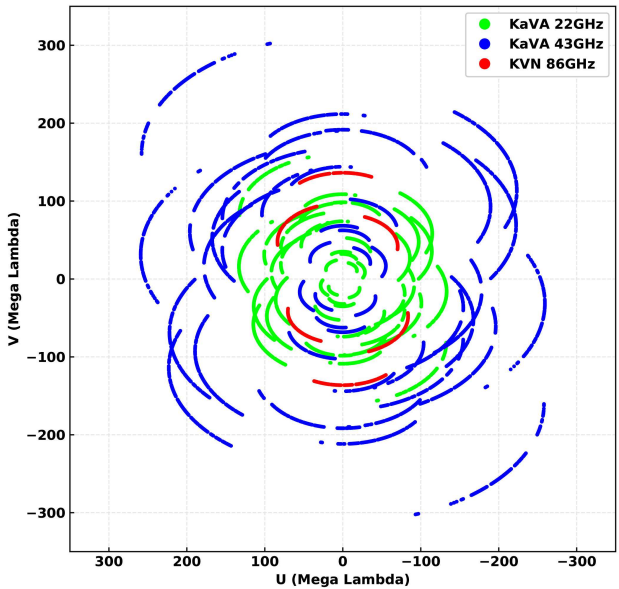}
 \caption{Representative $uv$-coverage of KaVA 22/43\,GHz and KVN 86\,GHz observations of \bo{} (in the unit of $10^6\lambda$ where $\lambda$ is the corresponding wavelength). The data are from the session in 2018 January.}
 \label{fig:uv}
\end{figure}

\begin{figure}[ttt]
 \centering
 \includegraphics[angle=0,width=1.0\columnwidth]{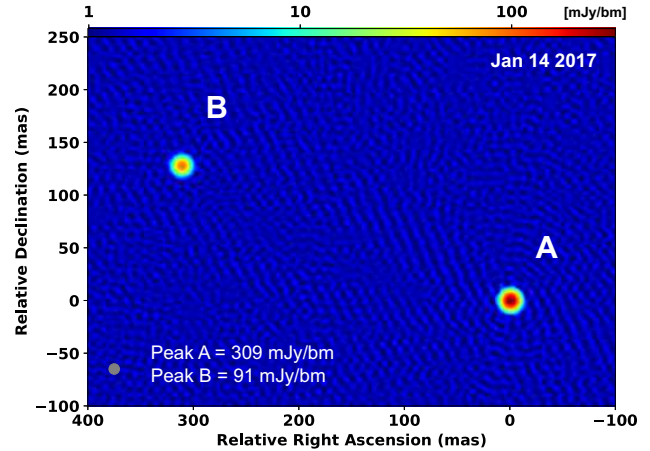}
 \caption{Wide-field KaVA 22\,GHz image of \bo{} that contains both of the lensed images A and B. The image from the 2017 January session is shown. The position of A is taken as the coordinate origin. The image is convolved with a 10 mas circular beam (gray circle at the bottom left corner of the image) to enhance the emission.}
 \label{fig:largescale}
\end{figure}

\begin{figure*}[bbb]
 \centering
 \includegraphics[angle=0,width=0.85\textwidth]{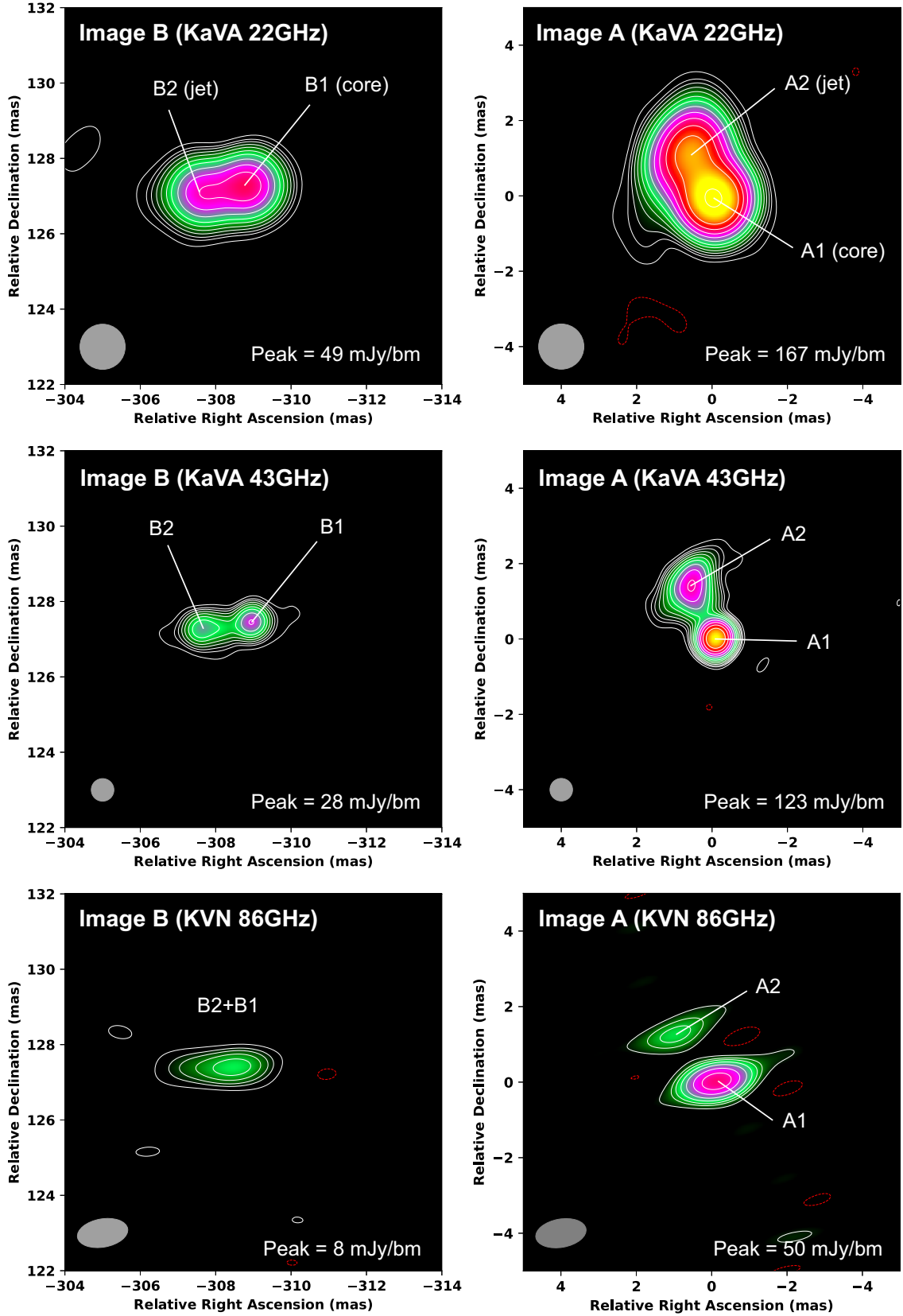}
 \caption{KaVA and KVN zoom-in images of A (left panels) and B (right panels), taken by KaVA at 22 GHz (top panels) and 43 GHz (middle panels) and KVA at 86 GHz (bottom panels). In each panel we show a stacked image over the four epochs.  
 For each image, contours start from $-1$ (red color), 1, 2,...times a 3$\sigma$ image rms level and increase by factors of $2^{1/2}$. The shape of the convolving beam is shown by the gray ellipse at the bottom left corner of each map.}
 \label{fig:stackimage}
\end{figure*}

\subsection{KVN 43 and 86\,GHz}

Truly in parallel to 5 out of the 9 KaVA 43\,GHz sessions, we also observed the source with the KVN-only array with 43\,GHz/86\,GHz dual-frequency simultaneous recording mode. These 5 dates are 2017 January 15 (MJD 57768), 2017 October 17 (MJD 58043), 2017 November 12 (MJD 58069), 2017 November 25 (MJD 58082), and 2018 January 5 (MJD 58123). For the session in 2017 November 25, only two stations were available due to an antenna motor issue at the Ulsan station, so we excluded this epoch from our study. A wideband 4\,Gbps mode was used where each frequency band was recorded at 2\,Gbps (a bandwidth of 512\,MHz for each band). We first calibrated the 43\,GHz data in the standard manner in AIPS. The derived 43\,GHz fringe phase solutions were then transferred to the 86\,GHz data by using the frequency-phase transfer (FPT) technique~\citep[e.g.,][]{rioja2011, algaba2015, zhao2018}. This greatly reduced the rapidly fluctuating phase components at 86\,GHz and allowed us to perform fringe fitting with a much longer solution interval than the typical coherence time at this frequency. Thanks to this strategy we are able to detect 86\,GHz fringes for the target at sufficient signal-to-noise ratios (S/Ns) for most of the scans. The KVN data were averaged in frequency and time in the same manner as for the KaVA data. Imaging was performed in Difmap. Since KVN is a three-element array, we performed phase-only self-calibration using a solution interval of 30\,s.

In Figure~\ref{fig:uv} we show the representative $uv$-coverage of our KaVA/KVN observations at each frequency. In Table~\ref{tab:observations} we summarize the basic information of individual KaVA/KVN images used in this paper. Typical angular resolution of KaVA (a maximum baseline length $D$$=$2300\,km) is 1.2\,mas (22\,GHz) and 0.6\,mas (43\,GHz), while that of KVN ($D$$=$560\,km) is 1\,mas at 86\,GHz. We assume the typical flux calibration accuracy of 10\% and 20\% for KaVA 22/43\,GHz and KVN 86\,GHz data, respectively.

\begin{figure}[ttt]
 \centering
 \includegraphics[angle=0,width=1.0\columnwidth]{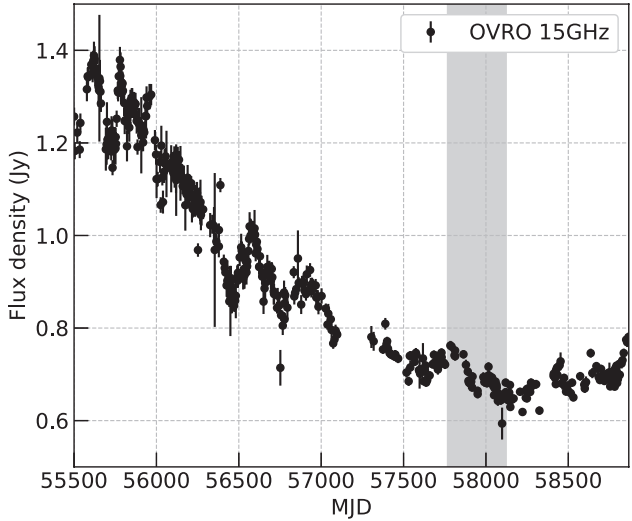}
 \caption{OVRO 15\,GHz long-term light curve of \bo. The shaded area indicates the span of our KaVA observations reported here. Note that OVRO (a single-dish radio telescope) measures total radio flux densities that integrate the emission from A, B, and the Einstein ring.}
 \label{fig:ovro}
\end{figure}

\section{Results}

\subsection{Lens plane properties at mm wavelengths}

In Figure~\ref{fig:largescale} we show a wide-field image of the lensed system obtained with KaVA at 22\,GHz. The lensed images A and B are clearly detected, while the diffuse Einstein ring emission is totally resolved out in our VLBI imaging due to the lack of short baselines sensitive to such a large-scale component. In Figure~\ref{fig:stackimage}, we show zoom-in mas-scale images toward A and B at 22, 43 and 86\,GHz, resolving the magnified pc-scale structures in both of the lensed images. These images are produced by stacking over the four sessions and better characterize the morphology at each frequency. At 22\,GHz the lensed image A shows a core-jet structure toward the north-northeast, while B exhibits a core-jet profile toward the east. 
The lensed image A additionally shows some diffuse extension in the direction perpendicular to the jet axis. These characteristics observed in our KaVA 22\,GHz images are in good agreement with those seen in the previous VLBA 22\,GHz images~\citep{spingola2016}. Hereafter we label A1 (core in A), A2 (jet component in A), B1 (core in B) and B2 (jet component in B), respectively.

On the other hand, the overall source brightness is significantly (a factor of 2) lower than the past levels. In Figure~\ref{fig:ovro}, we show a long-term 15\,GHz light curve of the source obtained by the Owens Valley Radio Observatory (OVRO) 40-m telescope~\citep{richards2011}. As can be seen, the historical radio flux of the target was gradually decreasing between MJD 55500 and MJD 58300, and the period of our KaVA observations was roughly coincident with a historical low state of the source, although some local fluctuations ($\sim$5\% in rms fractional amplitude) on time scales of months were seen.

\begin{figure}[ttt]
 \centering
 \includegraphics[angle=0,width=1.0\columnwidth]{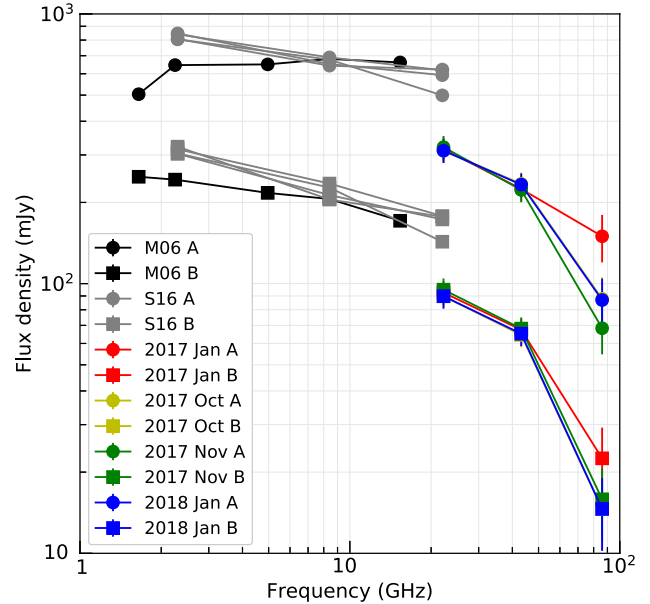}
 \caption{VLBI-scale integrated radio spectra of the lensed images A and B. The 22/43/86\,GHz data points colored in red, yellow, green, and blue are from the observations presented in this work. The low-frequency spectra colored in black and gray are taken from \citet{mittal2006} and \citet{spingola2016}, respectively.}
 \label{fig:spec}
\end{figure}

At 43 and 86\,GHz, the results presented here are the first robust VLBI detection and imaging of the lensed system \citep[but see also][]{porcas1996, lee2008}. In the KaVA 43\,GHz images the core-jet structure is more clearly resolved and each lensed image can be well characterized by a bright core and a secondary component separated by $\sim$1.5\,mas (in A) and $\sim$1.3\,mas (in B) from the core. At 86\,GHz, VLBI observations are generally challenging, but the unique capability of the KVN wideband multi-frequency receiving system allows us to detect and image the source at adequate S/N. For the lensed image A (typically S/N$>$30), the core-jet morphology toward the north-east is consistently seen up to 86\,GHz.  
The lensed image B is detected at S/N$\sim$10 in each epoch.  Its morphology is dominated by a single component, but there is a clear elongation toward the east, which is consistent with the structure seen at 22 and 43 GHz.

\begin{table}[ttt]
 \begin{minipage}[t]{1.0\columnwidth}
  \centering 
  \caption{Spatially resolved spectral indices\label{tab:spec}}
    \begin{tabular*}{1.0\columnwidth}{@{\extracolsep{\fill}}lcc}
    \hline
    \hline
     Feature & $\alpha_{\rm KQ}$\,(22--43\,GHz) & $\alpha_{\rm QW}$\,(43--86\,GHz)  \\
    \hline
     A (total) & $-0.48\pm0.05$ & $-1.28\pm0.42$ \\
     A1 (core) & $-0.45\pm0.17$ & $-0.76\pm0.30$ \\
     A2 (jet) & $-0.53\pm0.11$ & $-2.51\pm0.85$ \\
    \hline
     B (total) & $-0.49\pm0.05$ & $-2.00\pm0.25$ \\
     B1 (core) & $-0.32\pm0.15$ & -- \\
     B2 (jet) & $-0.53\pm0.10$ & --  \\
    \hline 
    \end{tabular*}
  \end{minipage}
  \tablecomments{Each spectral index is a mean value over the 4 epochs. The error of each spectral index is the standard deviation over the 4 epochs. $\alpha_{\rm QW}$ in B
  was estimated only for the total flux due to the low image S/N of B at 86\,GHz.}
\end{table}

\subsection{Radio spectra}

In Figure~\ref{fig:spec}, we show the observed integrated radio spectra of A and B up to 86\,GHz. For comparison, the low-frequency spectra from the literature~\citep{mittal2006, spingola2016} are also plotted. One can see that the (integrated) spectra for both A and B tend to become steeper with increasing frequency. Between 22 and 43\,GHz both A and B show very stable spectral shapes over the four epochs with an averaged spectral index of $\alpha_{\rm A, KQ}$$=$$-0.48\pm0.05$ or $\alpha_{\rm B, KQ}$$=$$-0.49\pm0.05$, respectively. At 86\,GHz the emission is quite variable and the spectral slope significantly changes with time. For the lensed image A, the integrated spectral index between 43 and 86\,GHz varies between $-0.58$ and $-1.70$ with a mean value of $\alpha_{\rm A, QW}$$\sim$$-1.28$. For the image B, $\alpha_{\rm B, QW}$ is generally steeper than $\alpha_{\rm B, KQ}$ with a mean slope of $\alpha_{\rm B, QW}$$\sim$$-2.0$. However, the image quality of B at 86\,GHz is rather limited and some amount of low-level flux (from the extended emission associated with the jet) in B may not be recovered at 86\,GHz. 

In Table~\ref{tab:spec}, we also show the spatially resolved spectral indices for the individual features A1, A2, B1, and B2. As typical for a blazar, the jet (A2 and B2) has steeper spectra than the core (A1 and B1). Nevertheless, it is notable that the core spectra themselves are substantially steep especially between 43 and 86\,GHz.

\subsection{Flux magnification ratio}
\label{sec:flux_magnification_ratio}

\begin{figure}[ttt]
 \centering
 \includegraphics[angle=0,width=1.0\columnwidth]{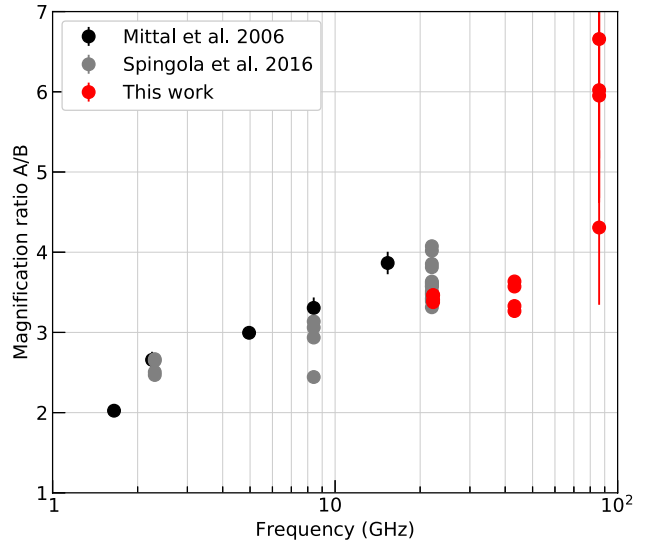}
 \caption{Magnification ratio A/B (as measured by VLBI) as a function of observing radio frequency. The 22/43/86\,GHz data points colored in red are obtained from the present observations. The uncertainties are estimated based on the S/N of recovered flux of each lensed image (while the systematic calibration uncertainty that is common to A and B is canceled out in the ratio).  The large errors for the 86\,GHz ratios are mainly caused by the low image S/N in image B, for which we assume $20$\% uncertainty in the recovered fluxes. The low-frequency data points colored in black and gray are taken from \citet{mittal2006} and \citet{spingola2016}, respectively.}
 \label{fig:ratio}
\end{figure}

In Figure~\ref{fig:ratio}, we plot the observed magnification ratio A/B (i.e., (A1+A2)/(B1+B2)) as a function of frequency. The flux ratio can be determined much more accurately than the individual fluxes since any systematic flux calibration errors for A or B are canceled out. 
To obtain a wider frequency coverage of A/B ratio, we also include the previous results obtained at 22\,GHz and lower frequencies~\citep{mittal2006, spingola2016}. The previous observations between 1.7 and 22\,GHz reported a gradual increase of the ratio with frequency, from $\sim$2 at 1.7\,GHz to $\sim$3.4--4 at 22\,GHz. Our magnification ratio measurements at 22\,GHz result in $\sim$3.4, which is in good agreement with the previous results. At 43\,GHz, the measured ratios are similar to the ones at 22\,GHz with a slightly larger scatter. At 86\,GHz, the scatter is still very large and the apparent large ratios are likely caused by the insufficient flux recovery of lensed image B, so the obtained ratios at 86\,GHz are less constrained than at 22/43\,GHz and should be considered as upper limits. 

Using the KaVA 43\,GHz images where the core and jet are well resolved, we also checked the magnification ratio for the core (A1/B1) and jet (A2/B2) separately. We find that A1/B1 tends to be systematically larger than A2/B2, such that A1/B1$=$3.3--4.5 (with a mean value 3.7) while A2/B2$=$3--3.3 (with a mean value 3.2).  We will discuss the possible reasons for the difference in Section~\ref{sec:lensmodel}.

\section{Discussion}
\subsection{Physical properties of the parsec-scale jet}
\label{sec:mm-site}

The gravitationally lensed blazar \bo{} provides a rare opportunity to study the physics of VHE $\gamma$-ray-emitting relativistic jets at $z$$\sim$1. High-resolution radio observations are especially useful to directly resolve the detailed structure of the innermost jet regions. Previous extensive VLBI imaging of this source ~\citep[e.g.,][]{biggs2003, mittal2006, spingola2016} revealed a core-jet morphology (in both A and B) at pc scales that is typical for powerful blazars. However, \citet{mittal2007} found that the low-frequency (1--15\,GHz) radio spectra of the lensed image A (and accordingly the A/B flux ratios) were significantly affected by the free-free absorption (FFA) in the lens galaxy, preventing us from imaging the pure GL morphology. In addition, (even though the external effects are irrelevant) the jet base could become opaque by the effect of synchrotron-self-absorption (SSA), which might make the access to the high-energy $\gamma$-ray emission site difficult at low frequencies~\citep[e.g.,][]{2009Sci...325..444A,spingola2016}.

The KaVA/KVN observations presented here have for the first time revealed the detailed mas-scale structures of this lensed system at mm wavelengths. In particular, we have found that the simultaneous mm-spectra of the radio core and jet become progressively steep with frequency. This indicates that the radio structure becomes substantially transparent to any absorption effects at these frequencies. In fact, if we adopt the best-fit FFA parameters derived by the low-frequency radio spectra toward A \citep[e.g., $T_{\rm e} = 10^4\,\rm{K}$ and $EM = 1.8\times 10^7 \rm{cm^{-6}\,pc}$ where $T_{\rm e}$ and $EM$ are the temperature and emission measure of the presumed H$_{\rm II}$ region in the lens galaxy;][]{mittal2007}, the FFA optical depth at $\geq$43\,GHz results in $\tau_{\rm FFA}\leq 2\times10^{-3}$, being essentially negligible. Therefore, we start to see the absorption-free radio structure of \bo{} at these frequencies. Note that we do not rule out the possible presence of a highly SSA-thick substructure in the core (i.e., a spectral turnover $\nu_{\rm SSA}>86$\,GHz). However, the contribution from such an inverted spectral component should be tiny at the observed frequencies.

It is notable that the mm-core spectra between 43 and 86\,GHz are quite variable, while those between 22 and 43\,GHz are rather stable over a year (Figure~\ref{fig:spec}). If we attribute the observed spectral steepening at the high frequencies to the radiative cooling of synchrotron electrons, we can estimate the magnetic field strength of the radio-emitting region through $B$$\sim$$1.3\,t_{\rm syn, obs}^{-2/3}\nu_{\rm b, obs}^{-1/3} \delta^{-1/3} (1+z)^{1/3}$ (Gauss), where $B$, $t_{\rm syn, obs}$ and $\nu_{\rm b, obs}$, $\delta$ are the observed time scale of synchrotron cooling (in yr), observed break frequency (in GHz) and Doppler factor, respectively~\citep[e.g.,][]{pacholczyk1970}. The 86\,GHz emission is clearly variable between 2017 January and October (0.75\,yr) and possibly even on shorter time scales of $\sim$1 month (2017 October--November). This suggests that $B$ of the 86\,GHz emission site to be at least $B$$\sim$$0.33\delta^{-1/3}$\,G (for $t_{\rm syn, obs}=0.75\,{\rm yr}$) and possibly as large as $\sim$$1.5\delta^{-1/3}$\,G (for $t_{\rm syn, obs}=0.08\,{\rm yr}$). The Doppler factor $\delta$ of the radio emission site is still not well constrained since the jet morphology is largely stationary, but if we assume that $\delta$$\sim$$20$ \citep[adopted in a SED model by][]{ahnen2016} is a maximum value, we can obtain $B$$\sim$0.12--0.55\,G.

Interestingly, the derived $B$-field strength seems to be somewhat larger than that implied for the $\gamma$-ray-emitting site in the previous spectral energy distribution (SED) modeling~\citep[0.03\,G;][]{ahnen2016}. In fact, they indicate that the broadband radio-to-$\gamma$-ray emission cannot be modeled by a simple single-zone model but can be better described by a two-zone model where the $\gamma$-ray emission region is offset from the low-energy emission region. 
It should be noted that the above-mentioned model also did not explain the emission observed from this source at $\lesssim 100$\,GHz. 
Our $B$-field estimate therefore seems to be consistent with this picture, implying that the primary mm emission region may originate in a more magnetized part of the jet.

Another remarkable feature seen in the data presented here is that the core-jet morphology in each lensed image remains extremely stable at least over two decades since the first VLBI images of this system were obtained~\citep{patnaik1995}.
As discussed in Section~\ref{sec:lensmodel}, the projected distance of the jet feature from the core is estimated to be $\sim$10\,pc in the source plane, implying a deprojected (viewing-angle-corrected) distance from the core to be of the order of $\sim$100\,pc. This feature is reminiscent of a ``recollimation shock" claimed in other relativistic jets exhibiting active $\gamma$-ray flares~\citep[e.g.,][]{cheung2007, marscher2008, hada2018}. If this is the case, the distant component could also be relevant to the $\gamma$-ray production. To test this hypothesis, it would be useful to cross-correlate the $\gamma$-ray light curves with spatially decomposed (core and jet) radio light curves.

\begin{table}[ttt]
 \begin{minipage}[t]{0.95\columnwidth}
  \centering 
  \caption{Position and flux densities of the lensed images at 43\,GHz used for lens modeling.\label{tab:lensed_images_properties}}
    \begin{tabular*}{0.9\columnwidth}{cccc}
\hline
\hline
 Lensed image  & $\Delta {\rm RA}$ & $\Delta {\rm Dec}$  & Flux density\\
& (mas) & (mas)  & (mJy) \\
\hline

 A1 & 0.0 & 0.0 & 129\\
 A2 & 0.7 & 1.3 & 106\\
 B1 & 309.1 & 127.4 & 35\\
 B2 & 310.4 & 127.2 & 33\\
\hline

    \end{tabular*}
     \end{minipage}
  \tablecomments{These values are the average over the three best data sets of KaVA observations (2017 October, 2017 November and 2018 January where the highest resolutions along jet axis are available). We conservatively assume uncertainties of 0.2~mas for the position and 10 \% for the flux densities.}
\end{table}

\begin{table}
 \begin{minipage}[t]{0.95\columnwidth}
  \centering 
  \caption{Recovered parameters from the lens mass modeling of \bo{}.\label{tab:lensmodel}}
    \begin{tabular*}{1.0\columnwidth}{@{\extracolsep{\fill}}ll}
\hline
\hline
Parameter & Value \\
 \hline
 ($\Delta$x, $\Delta$y)$_{\rm core}$ & $(0.1559, 0.0812) \pm  (0.0001, 0.0001)$ \\
($\Delta$x, $\Delta$y)$_{\rm jet}$ & $ (0.1569, 0.0821) \pm  (0.0001, 0.0001)$\\
 \hline
 $\Delta t_{\rm core}$ &  $11.6 \pm 0.02$\\
 $\Delta t_{\rm jet}$  & $11.5 \pm 0.02$\\
\hline

$\mu_{\rm A1}$, $\mu_{\rm B1}$ (core) &  $2.40\pm0.05$, $0.48 \pm 0.01$ \\
$\mu_{\rm A2}$, $\mu_{\rm B2}$ (jet) &  $2.42\pm 0.05$, $0.50\pm 0.01$\\
\hline
    \end{tabular*}
  \end{minipage}
  \tablecomments{Here we adopt the model of \citealt{wucknitz2004}, as described in Sec.~\ref{sec:lensmodel}. We optimize only for the position the source components at 43 GHz (core and jet), which are given in arcsec with respect to image A1, which is at (0,0) arcsec. We also provide the expected time delays between A1-B1 ($\Delta t_{\rm core}$) and A2-B2 ($\Delta t_{\rm jet}$) in days, where A is the leading image, and their magnification factors ($\mu$). We highlight that the uncertainties reported here do not take into account the uncertainty of the cosmological parameters. Therefore, they should be taken as lower limits of the uncertainties.}
\end{table}

\begin{figure*}
    \centering
    \includegraphics[width = 0.9\textwidth]{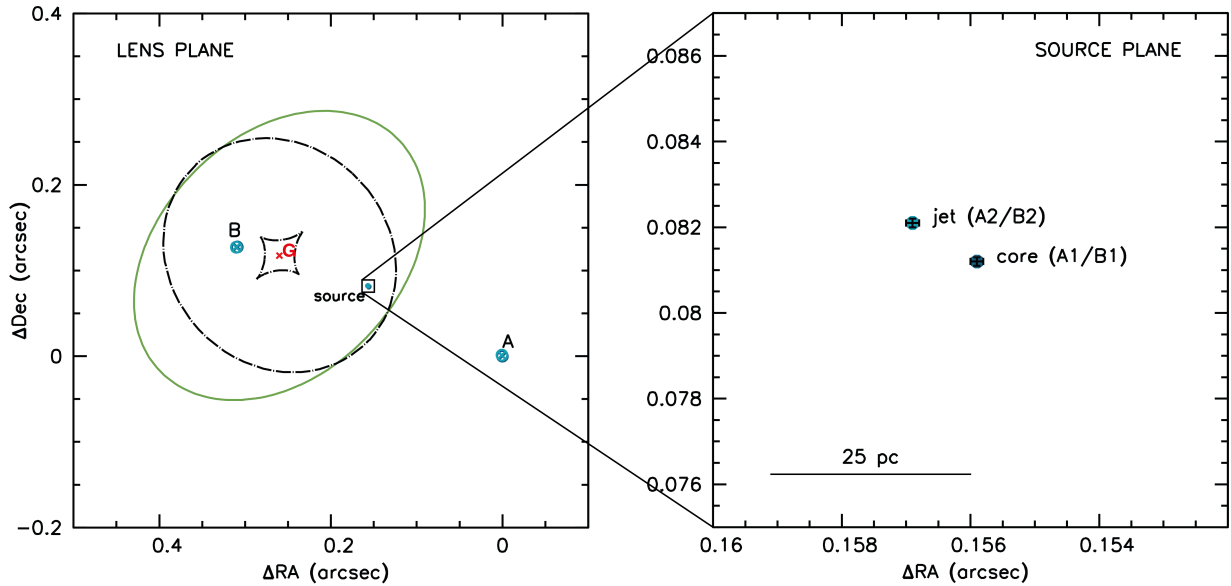}
    \caption{Left: lens mass model for \bo{}. We indicate the observed radio positions using open circles, while the model-predicted positions are represented by the crosses. All positions are given relative to component A1 (43\,GHz, Fig.~\ref{fig:stackimage}). The filled circles represent the position of the source components (core and jet). The lens critical curve is shown by the green solid line, while the source plane caustics are indicated by the dashed lines. The red cross indicates the position of the centroid of the isothermal elliptical power-law mass density distribution component. Right: zoom on the source plane. We indicate in dark blue the position of the core component (associated with the lensed images A1 and B1; see Fig.~\ref{fig:stackimage}), while the light-blue circle indicates the position of the jet (associated with the lensed images A2 and B2).}
    \label{fig:lensmodel}
\end{figure*}

\subsection{Lens modeling analysis}
\label{sec:lensmodel}

Having absorption-free VLBI images of \bo{}, now it would be instructive to apply for a simple lens model. This helps to discuss the intrinsic source geometry and also to check if the observed properties can be explained by a simple model or if any additional consideration is required. Here, in order to recover the intrinsic position (therefore, magnification) of the core and jet, we model \bo{} adopting a parametric lens modeling approach using {\sc gravlens} \citep{keeton2001}. We backward-ray trace the emission from this blazar by using average positions and flux ratios from the KaVA observations at 43\,GHz (those at the highest possible angular resolution; see Table~\ref{tab:lensed_images_properties}), because these data clearly resolves the subcomponents (see Fig.~\ref{fig:stackimage}, second row). We keep the lens mass model parameters fixed to the values obtained by \citet{wucknitz2004}, who inferred the gravitational potential at high precision using the entire extended emission from the lensed images and Einstein ring (see also \citealt{wucknitz2002, biggs2003}). In particular, we adopt their singular elliptical power-law model, without external shear,\footnote{The external shear in \bo{} has been found to be very small ($<$2\%; \citealt{lehar2000}), and was also not included in the models of \citet{wucknitz2002, wucknitz2004} and \citet{biggs2003}.} where the mass strength is $0.1616$ arcsec; the ellipticity and its position angle are $0.071$ and $-47$ degrees (east of north), respectively; the slope of the power-law is $-1.04$ and the lensing galaxy is at (0.2582, 0.1210) arcsec with respect to image A1, which is at (0,0) arcsec.
In order to estimate the uncertainty on the recovered parameters (source position, time delays and image magnifications), we use Monte Carlo simulations using the same methodology adopted by \citet{spingola2019} and  \citet{spingola2020a}. This method takes into account both the observational uncertainties on positions and fluxes (reported in Table~\ref{tab:lensed_images_properties}), and those of the lens mass model parameters \citep{wucknitz2004}. Lens and source plane obtained using this method are shown in Figure~\ref{fig:lensmodel} and the fitted parameters are summarized in Table\,\ref{tab:lensmodel}.

We find that (in the source plane) the core and jet \bo{} are separated by $1.4\pm0.2$~mas ($10.4\pm1.6$~pc) in projection, and they are located at about 110~mas from the center of mass of the lensing galaxy, as shown in Figure~\ref{fig:lensmodel}. The expected time delays between images A1-B1 and A2-B2 are found to be consistent with those measured by \citet{biggs2018} and \citet{cheung2014} within the uncertainties. As for the flux magnification, the lens model finds $\mu_{\rm A1} = 2.40\pm0.05$ and $\mu_{\rm B1} = 0.48\pm0.01$ for A1 and B1, respectively. Therefore, the model-predicted magnification ratio for the lensed images of the core results in $5$. This is broadly in agreement with the observed values (A1/B1$\sim$3.3--4.5; Section~\ref{sec:flux_magnification_ratio}), although a mean value in Table~\ref{tab:lensed_images_properties} is somewhat smaller than the predicated one. As for the jet part (lensed images A2 and B2), the model-predicted magnification ratio is $4.8$, consistent with that of the core. However, the observed ratios for the jet component (A2/B2$\sim$3--3.3) appear to be quite smaller than the predicted one. Therefore, the actual spatial gradient in the magnification ratio seems to be larger than predicted in the simple lens model applied here.

Given the absorption effects being negligible at our observing frequencies, an intriguing possibility for explaining the difference between predicted and observed magnification ratios is ``substructure lensing" by a compact clump in the lens galaxy (especially toward A). Such substructure lensing can locally affect individual parts of the background differently, leading to an additional magnification (or de-magnification) of the lensed source. For the lens geometry of \bo{}, a compact structure with projected size of the order of $\sim$10\,pc in the lens plane is enough to affect the core and jet components of a radio lensed image. Sizes of tens of pc are typical for a giant molecular cloud (GMC; see, e.g. \citealp{2020SSRv..216...50C}). Possible clumps in GMC can result in additional moderate amplifications by a factor of $1.5$ at perpendicular distance scales of parsecs (see \citealp{sitarek2016}). This would be able to explain the apparent discrepancy between the lens model predictions and observations.

Alternatively, we also note that since the image A is highly distorted by the lensing effect, the surface brightness in some stretched part of the jet (A2) could become either too faint to be detected by our VLBI sensitivity, or too extended to be sampled by the shortest baselines of the array. In this case we would recover only a fraction of the extended flux density that should be associated with A2, while the core (A1) should be unaffected because of its intrinsic compactness. 

Modeling the whole details of the observed lens properties is beyond the scope of this paper. Nevertheless, the discussion presented here indicates a good capability of high-frequency VLBI data for studying gravitational lensing at the smallest scales. A sophisticated lens modeling, including the absorption and substructure effects at the lens, together with deeper mm-VLBI images will allow us to better constrain the lensing potential and accordingly the physical properties of the innermost jet regions in distant AGN.  

\section{Summary}

We have reported the results on high-frequency KaVA/KVN observations of the $\gamma$-ray-emitting gravitationally lensed blazar \bo{}. We have successfully obtained the pc-scale images of the lensed system at 22, 43, and 86\,GHz. While the observed lensed morphology is overall consistent with that seen at low frequencies, the radio spectra become progressively steep at high frequencies, indicating that the innermost jet regions become optically thin to synchrotron emission and that the absorption effects 
due to the intervening lensing galaxy become negligible at millimeter wavelengths.  These results demonstrate that mm-VLBI observations can serve as a powerful tool for studying the intrinsic properties of lensed AGN, which may therefore better constrain the connection between radio and $\gamma$-ray emission. The KaVA/KVN monitoring presented here were performed along with a large MWL campaign. Extensive analysis on the broadband correlation and time-domain properties (e.g., jet kinematics and light curves) will be treated in forthcoming publications.

\acknowledgments 

We thank all KVN and VERA staff members who helped the operation. KVN is a facility operated by the Korea Astronomy and Space Science Institute. VERA is a facility operated by the National Astronomical Observatory of Japan in collaboration with associated universities in Japan. This work is supported by JSPS KAKENHI grant Nos. JJP18H03721 (K.N. and K.H.), JP19H01943 (K.H.), and JP18KK0090 (K.H.). C.S. is grateful for support from the National Research Council of Science and Technology, Korea (EU-16-001). This research has made use of data from the OVRO 40\,m monitoring program~\citep{richards2011} which is supported in part by NASA grants NNX08AW31G, NNX11A043G, and NNX14AQ89G and NSF grants AST-0808050 and AST-1109911.

\bibliographystyle{aasjournal}
\bibliography{kava_b0218}

\if0

\fi



\end{document}